\documentclass[%
 reprint,
superscriptaddress,
 amsmath,amssymb,
 aps,
floatfix,
]{revtex4-2}

\usepackage{graphicx}% Include figure files
\usepackage{dcolumn}% Align table columns on decimal point
\usepackage{bm}% bold math
\usepackage{hyperref}% add hypertext capabilities
\usepackage[utf8]{inputenc}
\usepackage[T1]{fontenc}
\usepackage{mathptmx}
\usepackage{gensymb}

\DeclareSymbolFont{epsilon}{OML}{cmm}{m}{it}
\DeclareMathSymbol{\epsilon}{\mathord}{epsilon}{"0F}

\DeclareUnicodeCharacter{2212}{-}

\begin{document}

\preprint{APS/123-QED}

\title{Fluids under tension: dynamics, mechanics, and structure of sparse active adhesive networks}

\author{Alex T. Grigas}
\affiliation{Department of Physics and BioInspired Institute, Syracuse University, Syracuse, New York 13244, USA}
\author{M. Lisa Manning}
\email{mmanning@syr.edu}
\affiliation{Department of Physics and BioInspired Institute, Syracuse University, Syracuse, New York 13244, USA}

%\date{\today}

\begin{abstract}
Sparse mechanical networks under tension are inherently prone to gelation/clustering instabilities as bonds break, so they typically cannot reconfigure without collapsing. Subverting this expectation, several biological materials have recently been found to maintain tension and uniformity while flowing. Here, we identify key ingredients needed to generate this fluid under tension and explore how details of the implementation affect material dynamics, structure, and mechanics. We find that activity must counteract the clustering instability, and that adhesive unbinding must be sufficiently stochastic so that it does not cancel out the activity and drive the system to a quiescent absorbing state. We find that the nature of the activity dramatically changes the material properties and glassy dynamics: self-propelled crawling, implemented as a force monopole, drives collective motion over long wavelengths, resulting in anti-hyperuniform density, whereas a force dipole representing particles pulling on each other drives scale-free collective motion and more uniform density. Both models exhibit exotic mechanics: stresses in each bond relax only through bond turnover, but the global tension never relaxes. In addition, in the crawling model the structure of the network relaxes faster than the pattern of stresses, suggesting the material retains a memory of its stress state even as the particle positions reconfigure. Together, this work shows that we can distinguish between activity mechanisms in biological systems by analyzing emergent materials properties, and design exotic emergent properties in reconfigurable active matter systems.
\end{abstract}

\maketitle

%\tableofcontents

\section{\label{sec:intro} Introduction}

\begin{figure*}[t]
\centering
\includegraphics{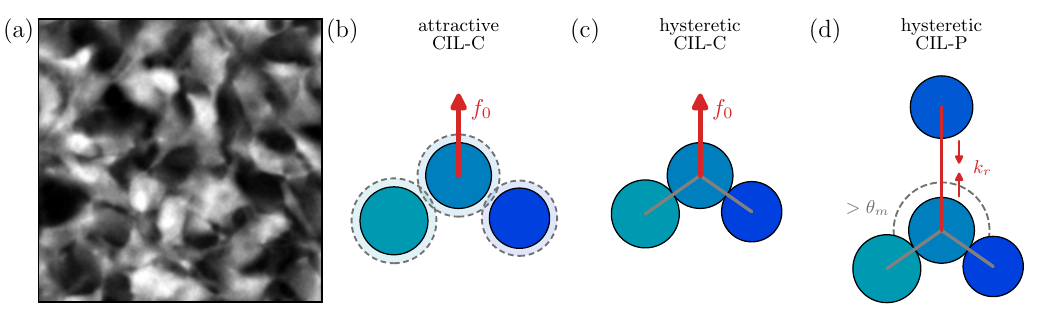}
\caption{\textbf{Different models for adhesive active particle networks} (a) An example of the sparse cellular network of the avian PSM cytoplasmic labeling, which is the inspiration for these models, see Ref~\cite{fut:Grigasbiorxiv2025} for more details. (b) The attractive particle model directing its activity away from its neighbors, CIL-C (red arrow). Repulsive cores are shown as solid disks, with the surrounding attractive shell in cyan. (c) The hysteretic particle model, which forms adhesions upon repulsive contact, under CIL-C activity (red arrow). (d) The hysteretic particle model under CIL-P activity, which reaches toward open space, binds a neighbor when the open angle exceeds $\theta_m$, and pulls it in at rate $k_r$ (red line and arrows).}
\label{fig:models}
\end{figure*}

Tension is often a sign of rigidity in materials. However, there are numerous examples of biological materials that appear to maintain tension while flowing and rearranging like a fluid. For example, it was recently shown that the avian presomitic mesoderm (PSM), a sparse cellular network of stellate cells, both retracts elastically following ablation, while flowing and rearranging during elongation~\cite{fut:Grigasbiorxiv2025}. (See Fig.~\ref{fig:models}(a)). At smaller scales, there is evidence that actin networks are also under tension and flow without collapsing~\cite{networks:WangPRXLife2026}.  

This is counterintuitive because in sparse networks like colloidal gels~\cite{gels:ZaccarelliJPhysCondMat2007,gels:LuNature2008,gels:ZiaJRheology2014}, tension leads to a clustering instability; when an existing bond releases and allows configurational rearrangements, the remaining tensioned bonds will shorten to lower their energy, opening a hole in the material. In some cases this clustering instability may be useful; while the avian PSM appears to maintain both tension and flow, others tissues appear to use similar instabilities to form structures such as somite condensation, intestinal villi and dermal follicles~\cite{mesenchyme:PalmquistCell2022,mesenchyme:HuyckeCell2024}.

Therefore, key open questions include: how can a tensioned network avoid the clustering instability and still be reconfigurable? And what dynamic, structural and mechanical properties does a tensioned fluid network possess? Are such features universal, or do they depend on model details? 

Previously, we showed that for a non-confluent material -- where there are large gaps between cells -- to maintain tension and avoid clustering while flowing, energy must be injected into the system in order to generate new tension and contacts. In addition, these active forces must be biased to oppose the clustering instability, i.e. away from one's neighbors and towards empty space~\cite{fut:Grigasbiorxiv2025}.

In biology, there are credible mechanisms that generate this type of activity in response to local density. Cell biologists have demonstrated that some cell types move away from neighbors upon contact, termed contact inhibition of locomotion (CIL)~\cite{CIL:Carmona-FontaineNature2008,CIL:DavisDevelopment2012,CIL:BischoffJCB2025}. Similarly, it has been recently proposed that actin networks~\cite{networks:WangPRXLife2026} preferentially create new network bonds in regions of low network density. It is intriguing to consider whether similar mechanisms could be programmed into active colloids~\cite{abp:BechingerRevModPhys2016} or active self-assembled DNA structures~\cite{activity:PatinoJACS2024,activity:MurphyPLoSONE2025} to drive similar emergent behavior. We term all active forces that drive particles away from regions of high neighbor density as ``CIL-like activity''.

A challenge to using these ideas for design and biological insight is that there are multiple ways for particles/cells/fibers to bias tension and contact generation towards empty space. An obvious choice that follows from active matter literature is based on self-propelled particles; we can simply direct a particle's active self-propulsion or crawling force away from nearby neighbors, which we term CIL-C~\cite{abp:RomanczukEPhysJ2012,abp:CatesAnnRevCondMatPhys2015,CIL:SmeetPNAS2016,abp:BechingerRevModPhys2016}. (See Fig.~\ref{fig:models} (b) and (c)). Active crawling models are inherently non-Hamiltonian as the crawling force is non-reciprocal; each cell is pushed along its activity direction according to its own local environment. Activity can also be generated using a pairwise force: inspired by observations that PSM cells develop actin-rich appendages that adhere to neighbors and then appear to contract/reduce the distance between neighbors, we developed a ``pulling" model we term CIL-P. (See Fig.~\ref{fig:models} (d) and Sec.~\ref{sec:methods}). Previously, we showed that both forms of activity can generate a uniform, tensioned network that can diffusively rearrange~\cite{fut:Grigasbiorxiv2025}. But are there identifiable differences in their emergent behavior and geometry that we could use to distinguish between them? More broadly, what does it take to build a fluid under tension, and what kind of material results? 

In this manuscript, we first show that the simplest model for adhesive particle interactions combined with CIL-like behavior exhibits a pathology: an absorbing state with no flow. We avoid the absorbing state via a hysteretic interaction potential with stochastic particle-particle unbinding at a rate $k_\mathrm{off}$. As rearrangements are initiated by a stochastic unbinding event, one might expect the bond lifetime $\tau_b = 1/k_\mathrm{off}$ to control the dynamics, structure, and mechanics. In all CIL models, we find the bond lifetime does set the timescale for stress relaxation, but in CIL-C models the structure relaxes faster due to collective excitations. As a consequence, CIL-C models retain a memory of their stress state even as the structure reconfigures, which is not the case in CIL-P.  This difference also leads to differences in large-scale structure between the two models, highlighting signatures that could distinguish between the two in experiments.

The remainder of the article is organized as follows. In Sec.~\ref{sec:methods}, we outline the numerical details of the models. In Sec.~\ref{sec:results} we present numerical results for simulations of three models for particle interactions and activity that generate a fluid under tension and describe how model details lead to distinct dynamical, mechanical and structural properties. In Sec.~\ref{sec:conclusions}, we present conclusions and future directions.

\section{\label{sec:methods}Methods}

\begin{figure}[t]
\centering
\includegraphics{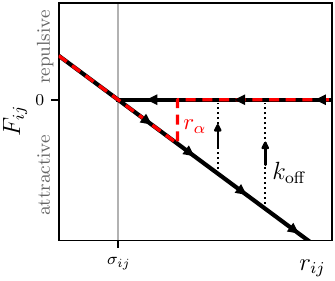}
\caption{\textbf{Comparison of attractive and hysteretic particle interactions.} The pairwise forces for the attractive particle model (red dashed line) and the hysteretic sticking model (black lines). The black arrows indicate the hysteresis, as particles first have to come into repulsive contact before forming a double-sided spring. Additionally, the bond breaks and returns to zero interaction according to a rate $k_{\rm off}$, i.e., with a stochastic adhesive range. The attractive interaction always extends to precisely $r_{\alpha}$.}
\label{fig:forces}
\end{figure}

To simulate a fluid under tension, we have to select two key components: the particle-particle interaction and the activity. Each particle follows overdamped dynamics at zero temperature,
\begin{equation}
    \gamma\,\dot{\mathbf{r}}_i = \sum_j \mathbf{F}_{ij} + \mathbf{F}_i^{\rm act},
\end{equation}
where $\mathbf{F}_{ij}$ is the interaction force between particles $i$ and $j$ and $\mathbf{F}_i^{\rm act}$ is the active force. The simplest way to add tension to the repulsive particle model, in which particles interact only on overlap, is to allow adhesive interactions that extend the interaction beyond contact,
\begin{equation}
\label{eq:attractive}
    \mathbf{F}_{ij} = k(\sigma_{ij} - r_{ij})\,\Theta(r_\alpha - r_{ij})\,\hat{\mathbf{r}}_{ij},
\end{equation}
where $r_{ij} = |\mathbf{r}_i - \mathbf{r}_j|$ is the interparticle distance, $\hat{\mathbf{r}}_{ij} = (\mathbf{r}_i - \mathbf{r}_j)/r_{ij}$, $\sigma_{ij} = (\sigma_i + \sigma_j)/2$ is the average particle diameter, which sets the repulsive contact distance, $\Theta(x)$ is the Heaviside step function, and $r_\alpha = \lambda\sigma_{ij}$ with $\lambda > 1$ is the attractive contact distance~\cite{attractive:LoisPRL2008,sticky:KoezePRL2018,attractive:GrigasPRE2024}. We refer to this as the attractive particle model (red dashed line in Fig.~\ref{fig:forces}). 

A similar attractive shell has been proposed for loosely packed yet cohesive tissues such as embryonic mesenchyme, in which cells are separated by wide matrix-filled gaps: a pericellular matrix rich in hyaluronic acid (HA) surrounds each cell, sets its spacing, and mediates adhesion while moving with the cell as the tissue rearranges~\cite{attractive:WinklbauerJCellSci2019}. In sparse, stellate mesenchyme such as the avian PSM, cells visibly hold on to each other through protrusions, and HA, which surrounds the cells and fills the intercellular gaps, alters the mechanics of the tissue~\cite{mesenchyme:MichautCurrBio2025}. HA is likely playing multiple roles: (1) contributing, alongside the protrusions, to tensile attractions between cells, (2) generating a swelling force in the intercellular space, and (3) adding a repulsive layer on the cell surface. The attractive shell of Eq.~\eqref{eq:attractive} captures (1) and the repulsive core captures (3). It cannot capture (2), since the same shell that fills the interparticle gaps must also pull cells together and cannot represent a swelling pressure.

Therefore, we also consider a hysteretic sticking potential, which we introduced in Ref.~\cite{fut:Grigasbiorxiv2025} to model the avian PSM, shown as black lines in Fig.~\ref{fig:forces}. To make the interaction hysteretic, we give each pair a bond variable $b_{ij} \in \{0, 1\}$. Particles bind when they come into repulsive contact, and a bonded pair is connected by a double-sided spring,
\begin{equation}
    \mathbf{F}_{ij} = k(\sigma_{ij} - r_{ij})\,b_{ij}\,\hat{\mathbf{r}}_{ij},
    \label{eq:hysteretic}
\end{equation}
which repels on overlap and attracts on extension. Each bond unbinds stochastically, with its lifetime $t$ drawn from the exponential distribution
\begin{equation}
    P(t) = k_{\rm off}\,e^{-k_{\rm off} t},
\end{equation}
so that bonds unbind at rate $k_{\rm off}$ and live for an  average time $\tau_b = 1/k_{\rm off}$. At a given separation $r_{ij} > \sigma_{ij}$, a pair is therefore attracted or not depending on whether it has touched since its last unbinding, which gives the hysteresis loop in Fig.~\ref{fig:forces}. A pair that unbinds while overlapping rebinds immediately, so overlapping particles always repel. 

Here we use purely stochastic unbinding with no force dependence to serve as a neutral baseline, but future work could easily extend this to capture catch and slip bonds, respectively, where the rate of unbinding decreases or increases with increasing force~\cite{catchbonds:ThomasAnnRevBiophys2008,catchbonds:RakshitPNAS2008}. Additionally, future extensions of the hysteretic model could capture HA swelling forces; for example, adding an area force on the enclosed polygons formed by the double-sided bonds.

We now consider the form the $\mathbf{F}_i^{\rm act}$ for capturing CIL-like behavior. In standard self-propelled particles, each cell carries an active force of fixed magnitude $f_0$, with an orientation that diffuses over some persistence time. Instead we specify that the orientation of the force always points away from bonded neighboring cells,
\begin{equation}
    \mathbf{F}_i^{\rm act} = f_0\,\hat{\mathbf{e}}_i, \qquad
    \hat{\mathbf{e}}_i = \frac{\mathbf{m}_i}{|\mathbf{m}_i|}, \qquad
    \mathbf{m}_i = \sum_j c_{ij}\,\hat{\mathbf{r}}_{ij},
    \label{eq:cilc}
\end{equation}
where $c_{ij} = 1$ if cells $i$ and $j$ are interacting and 0 otherwise: $c_{ij} = \Theta(r_\alpha - r_{ij})$ in the attractive model and $c_{ij} = b_{ij}$ in the hysteretic model. (See Fig.~\ref{fig:models} (b) and (c)). A cell with no contacts has no active force. Here, we consider the limits in which the CIL repolarization is instantaneous and there is no noise on the active direction, although such extensions would be interesting avenues for future work. As this activity does not conserve momentum and mimics cells crawling on a substrate, we refer to it as the CIL-C activity model.

\begin{figure*}[t]
\centering
\includegraphics{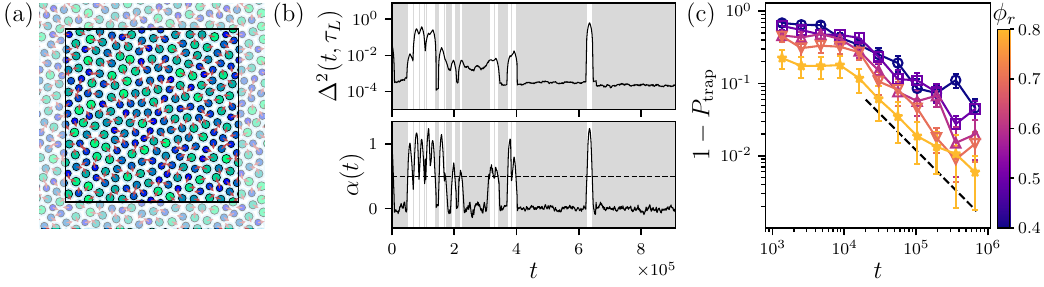}
\caption{\textbf{Attractive particles under CIL-C find an absorbing state.} (a) An example configuration of the attractive particle model. Particles are colored by size increasing from blue to green. Each particle as an attractive shell (cyan) and moves under CIL-C active (red arrow). Periodic images are shown with transparency. (b) The mean-squared displacement $\Delta^2(t,\tau_L)$ at short lag-time $\tau_L$ and the slope $\alpha$ of $\Delta^2(t)$ over the same window over time $t$. Grey shading indicates trapped states where $\alpha<1/2$ (grey dotted line). (c) $1-P_{\rm trap}$ vs $t$ colored by repulsive packing fraction $\phi_r$ increasing from purple to orange. Dashed line indicates a slope of $-1$.}
\label{fig:absorbing}
\end{figure*}

Another type of activity is inspired by observations of mesenchymal cell-cell adhesion remodeling. We previously proposed that particles could also reach out away from their current neighbors and towards empty space, making adhesive contacts with new neighbors and pulling them in~\cite{fut:Grigasbiorxiv2025}. We term this the CIL-P model, which is as follows: cells carry no active force, $\mathbf{F}_i^{\rm act} = 0$; instead, the activity enters through the bonds. Each bond is a double-sided spring with rest length $\ell_{ij}$,
\begin{equation}
    \mathbf{F}_{ij} = k\,(\ell_{ij} - r_{ij})\,b_{ij}\,\hat{\mathbf{r}}_{ij},
    \label{eq:cilp}
\end{equation}
and bonds form in two ways. As in the hysteretic model, two cells that come into contact bind with $\ell_{ij} = \sigma_{ij}$. In addition, unbonded cells that are neighbors in the radical Voronoi tessellation bind at rate $k_{\rm on}$, at any distance, provided the new bond points into an open gap on both cells: if $\theta_i$ is the angle between the two bonds of cell $i$ on either side of the direction to $j$ ($2\pi$ if $i$ has fewer than two bonds), and $\theta_j$ likewise for cell $j$, binding requires $\theta_i > \theta_m$ and $\theta_j > \theta_m$, depicted in Fig.~\ref{fig:models} (d). A bond formed this way starts unstressed, with $\ell_{ij}$ equal to the separation $r_{ij}(t_0)$ at its formation time $t_0$, and its rest length then contracts linearly until it reaches contact,
\begin{equation}
    \ell_{ij}(t) = \max\!\left[\,r_{ij}(t_0) - k_r\,\sigma_{ij}\,(t - t_0),\; \sigma_{ij}\right],
    \label{eq:ratchet}
\end{equation}
so that the bond pulls the two cells together at a rate set by $k_r$. All bonds unbind with the same exponential lifetime distribution, at rate $k_{\rm off}$, similar to the CIL-C model.

This pulling mechanism relies on the hysteresis: because a bond is a state of the pair rather than a function of its separation, it can form at any distance and carry its own rest length. In the attractive particle model, the force is fixed by the separation alone, so a pair beyond $r_\alpha$ cannot hold on to each other and there is no bond whose rest length could contract. Like $k_{\rm off}$, we model $k_{\rm on}$ as purely stochastic. Adding further feedback, such as responding to separation, would be an interesting avenue for future work.

Bond turnover and binding are integrated as continuous-time stochastic processes on the time-step grid. Each bond draws an exponentially distributed lifetime with mean $1/k_{\rm off}$ when it forms and is removed at the first time step after it expires, which gives the same survival probability as testing every bond at every step. In CIL-P, binding attempts occur as a Poisson process at total rate $k_{\rm on}$ times the number of Voronoi neighbor pairs; each attempt selects a random pair and binds it only if both cells have an open angle exceeding $\theta_m$ in its direction, so that every qualifying pair binds at rate $k_{\rm on}$. The radical Voronoi tessellation is recomputed every 500 steps, short compared to $1/k_{\rm on}$ and $1/k_r$, and the rest length of each ratcheting bond is decreased by $k_r\sigma_{ij}\Delta t$ every time step until it reaches $\sigma_{ij}$. Initial positions are drawn at random at $\phi = 0.4$ and relaxed to a force-balanced state by FIRE at fixed density~\cite{fire:BitzekPRL2006}. Each system then evolves for $5\tau_b$ before measurements begin, and all step strains are applied to this equilibrated state. System sizes of $N=64-4096$ are simulated throughout, and all simulations are under periodic boundary conditions.

We report all quantities in simulation units. To prevent crystallization, diameters are chosen uniformly from $[\sigma_{\rm min},\sigma_{\rm max}]$ where $\sigma_{\rm min}=1.0$ and $\sigma_{\rm max}=1.4$. Lengths are measured in units of $\sigma_{\rm min}$, forces in units of $k\sigma_{\rm min}$, and time in units of $\gamma/k$, the relaxation time of a single spring. Rates ($k_{\rm off}$, $k_{\rm on}$, $k_r$) are therefore in units of $k/\gamma$, the crawling force $f_0$ in units of $k\sigma_{\rm min}$ and stresses, the tension $\mathcal{T}$ and the elastic moduli in units of $k$. The equation of motion is integrated with a forward Euler scheme with time step $\Delta t$. For the attractive CIL-C model, $\Delta t$ is varied, since its dynamics depend on $\Delta t$. (See Appendix~\ref{appendix:dt}). For the hysteretic models, $\Delta t = 0.02$ (CIL-C) and $\Delta t = 0.01$ (CIL-P) were chosen so that measured quantities change by only a few percent on further reducing $\Delta t$; CIL-P requires the smaller step because its structural relaxation time is more sensitive to $\Delta t$. 

\section{\label{sec:results}Results}

Here we present the numerical results for all three models of a fluid under tension. We first show in Sec.~\ref{sec:sticky} how the attractive particle model gets trapped in an absorbing state under CIL-C activity where the activity balances the attractions trapping the system. Therefore, for the remainder of the manuscript we focus on a hysteretic sticking model that avoids the absorbing state due to its stochastic unbinding. In Sec.~\ref{sec:dynamics} we present dynamical and finite-size analysis of both CIL-C and CIL-P activity models and in Sec.~\ref{sec:structure} we examine both models' long-range structure. In Sec.~\ref{sec:mechanics}, we measure both the shear and bulk response.  Lastly, in Sec.~\ref{sec:memory} we compare and contrast CIL-C and CIL-P's relaxation of stress and configurational memory.

\subsection{\label{sec:sticky}The attractive particle model under CIL-C activity generates a non-fluid absorbing state}

To be in the tensioned fluid phase under CIL-C, the crawling force must be large enough for particles to crawl away from their neighbors and make new contacts~\cite{fut:Grigasbiorxiv2025}. We begin by studying the attractive particle model at the low repulsive packing fraction $\phi_r=0.4$. The attractive range $r_\alpha$ defines the attractive packing fraction $\phi_a$. With a short-range attraction and a strong enough crawling force $f_0$, the initially randomly placed particles will push away from their neighbors, until eventually settling in an absorbing state where activity goes to zero with no attractive overlaps, in a similar manner to the absorbing state in random/biased--organization models (RO/BRO)~\cite{bro:CorteNatPhys2008,bro:WilkenPRL2021,bro:GallianoPRE2026}. When $\phi_a$ is low and $f_0$ is stronger than the attractions, CIL-C is nearly identical to BRO.

As $\phi_a$ is increased towards the jamming density near $\phi\sim0.82$, a tensioned fluid-like phase emerges where the system initially maintains activity, is highly structurally uniform and bears tension. (See Fig.~\ref{fig:absorbing} (a)). However, we find that while nearly every particle is active for high $\phi_a=0.9$, the system can become globally arrested. We measure the trapping using the mean-square displacement,
\begin{equation}
\Delta^2(t,\tau) = \left\langle |\mathbf{r}_i(t+\tau) - \mathbf{r}_i(t)|^2 \right\rangle_i ,
\end{equation}
with the center-of-mass motion subtracted. To follow trapping in time, we average $\Delta^2(t',\tau)$ over all time origins $t'$ in $[t, t + 2\tau_L]$ and define its local exponent $\alpha(t) = d\ln\Delta^2/d\ln\tau$ over $\tau_L/10 \le \tau \le \tau_L$; a window is trapped when $\alpha(t) < 1/2$. We show example time traces of $\Delta^2(t,\tau_L)$ and $\alpha(t)$ in Fig.~\ref{fig:absorbing} (b). While the system flows at the start of the simulation, it quickly falls into long-lived trapped states with $\alpha\sim0$. The example traces also show brief intermittency of trapping.

\begin{figure*}[t]
\centering
\includegraphics{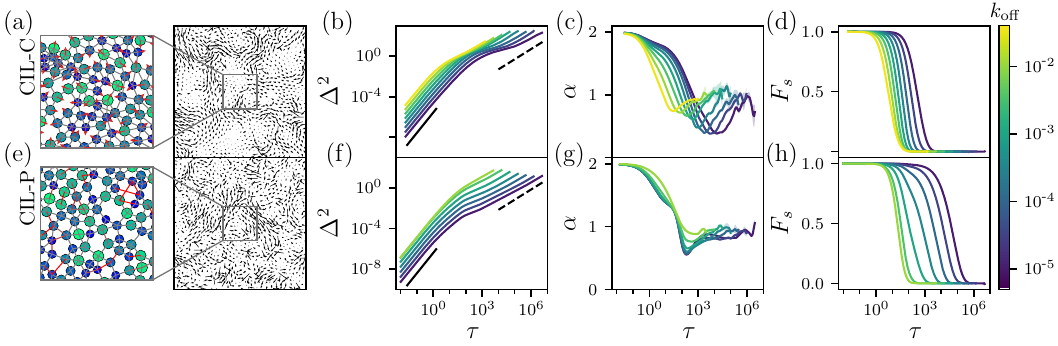}
\caption{\textbf{Dynamics of crawling (top row) and pulling (bottom row) differ.} (a,e) Displacement field over one relaxation time with zoom showing CIL-C (red arrows) and CIL-P (red bonds) (b,f) The mean-squared-displacement $\Delta^2$ vs lag-time $\tau$. (c,g) The local slope extracted from $\Delta^2$ vs $\tau$. (d,h) The intermediate scattering function $F_s$ vs $\tau$ measured at the wave vector corresponding to the nearest neighbor peak. $N=1024$.}
\label{fig:dynamics}
\end{figure*}

Does the attractive system approach a true absorbing state with no global motion or will it continue to jump from trap to trap? In Fig.~\ref{fig:absorbing} (c) we show the ensemble averaged probability for the system to be trapped approaching unity via $1-P_{\rm trap}$ over time for $\phi_r=0.4$, $\phi_a=0.9$, $f_0=1.5$ and $N=256$. We find $P_{\rm trap}$ to start low and continually increase over time. Additionally, we find that if $\phi_r$ is increased, while keeping $\phi_a=0.9$ and adjusting $f_0$ so that the ratio of attractive force to crawling force is constant, that $P_{\rm trap}$ approaches $1$ faster. We confirm these results as a function of the time step $\Delta t$ in Appendix~\ref{appendix:dt}. The results in Fig.~\ref{fig:absorbing} are in the converged small time step limit. 

To generate a fluid under tension, the activity needs to fight the attractive force in the particle-particle interaction. However, we find that if both the activity and attractions are deterministic functions of the particle separation with no memory or time delay, there exist states where they are able to exactly cancel each other, leading to an absorbing or dynamically arrested state. When such states exist, the system generically finds them, for the same reason as in random organization models. These ``pathological'' frozen states do not capture the fluid behavior we seek to explain, though they may be of interest to statistical physicists as another example of random organization.

In contrast, because the hysteretic model has stochasticity in its unbinding, there are random time delays in the adhesive forces that prevent them from canceling out active forces over extended periods of time, and we never find a pathological absorbing state. Therefore, the remainder of this manuscript focuses on the hysteretic sticking model. 

\subsection{\label{sec:dynamics}Hysteretic Particle Dynamics under CIL-C and CIL-P}

The hysteretic sticking model results in an underlying elastic network of cells whose remodeling is explicitly controlled by the unbinding time $\tau_b=1/k_{\rm off}$. Does this mean that both models' structural memory is also directly controlled by $\tau_b$? Moreover, while both crawling and pulling models generate a fluid under tension through a similar CIL-like mechanism, are its emergent features universal, or can we distinguish between them? In Fig.~\ref{fig:dynamics} we present displacement fields of (a) CIL-C and (e) CIL-P with a zoom of the underlying particles for $N=1024$. Below, we analyze the differences seen in these displacement fields over all time origins to quantify the important differences and their origins in the two different forms of activity.

\begin{figure*}[t]
\centering
\includegraphics{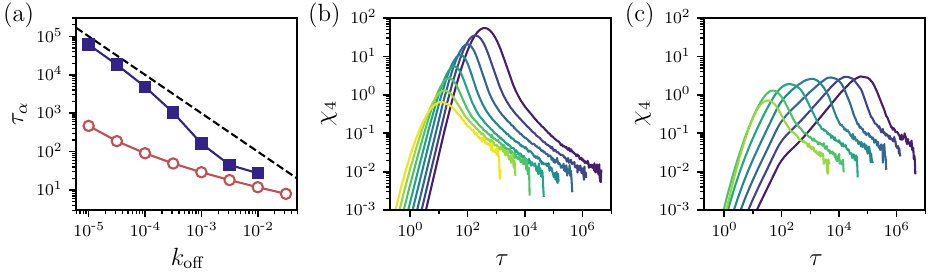}
\caption{\textbf{Structural memory is governed by the scale of collective dynamics.} (a) $F_s$ relaxation time $\tau_{\alpha}$ vs the unbinding time $k_{\rm off}$ for CIL-C (red circles) and CIL-P (blue squares). The black dashed line indicates $\tau_{\alpha}=1/k_{\rm off}=\tau_b$. The four-point susceptibility $\chi_4$ vs lag-time $\tau$ for (b) CIL-C and (c) CIL-P, colormap shows decreasing $k_{\rm off}$ from yellow to purple, $N=1024$. The structural timescale is the same as the unbinding time in CIL-P where the scale of collective dynamics remains small, but it is much shorter in CIL-C where there is a growing length scale for collective dynamics.}
\label{fig:tau_chi}
\end{figure*}

To begin understanding the dynamics of both systems, we start by analyzing $\Delta^2$ over time for small systems of $N=1024$ over a wide range of unbinding rates $k_{\rm off} = 10^{-2}-10^{-5}$, inside the tensioned fluid regime with a high crawling force $f_0=1.0$ for CIL-C. Similarly, to be within the tensioned fluid regime, we simulate CIL-P with $k_{\rm on}=0.02$, $k_r=0.02$ and $\theta_m=2~\rm rad$. As $k_{\rm off} \rightarrow k_{\rm on}$, CIL-P begins to fall out of the tensioned fluid phase. All hysteretic particle simulations were conducted at $\phi=0.4$ to avoid any complications from possible cage-controlled glassiness. In Fig.~\ref{fig:dynamics} (b,f) we plot $\Delta^2$ over time lag $\tau$ versus $k_{\rm off}$ for CIL-C and CIL-P respectively. Both models slow down as $k_{\rm off}$ decreases, and transition from a ballistic regime to diffusive regime. We extract the local slope of $\Delta^2$ versus $\tau$ and find distinct ballistic-diffusive crossover behavior for both models. (See Fig.~\ref{fig:dynamics} (c,g)). For CIL-C, as $k_{\rm off}$ decreases, the extent of subdiffusion increases from $\alpha\sim0.75-0.3$ and the subdiffusive onset time increases to longer times, on the order of $\tau_b$, before reaching the diffusive regime. In contrast, CIL-P reaches its subdiffusive point at the same time regardless of $k_{\rm off}$ with the same magnitude $\alpha\sim0.5$ before reaching a diffusive regime. 

We further quantify the dynamical relaxation using the self-intermediate scattering function~\cite{isf:KobPRE1995},
\begin{equation}
F_s(q,\tau) = \left\langle e^{i\mathbf{q}\cdot[\mathbf{r}_j(t+\tau) - \mathbf{r}_j(t)]} \right\rangle_{j,t,\hat{\mathbf{q}}},
\end{equation}
where the average runs over particles $j$, time origins $t$, and directions $\hat{\mathbf{q}}$ of the wave vector at fixed magnitude $q = 2\pi\sqrt{N}/L$, the wavenumber of the typical interparticle spacing $L/\sqrt{N}$. $F_s$ decays to 0 as particles move away from their original positions by more than one interparticle spacing. We plot $F_s$ versus $k_{\rm off}$ for both models in Fig.~\ref{fig:dynamics} (d, h). In both models $F_s$ decays in a single step, with no plateau, and is close to a single exponential (deviations $\lesssim 5\%$). Unlike in glass-forming liquids, where a subdiffusive MSD corresponds to a plateau in $F_s$ indicating caging, here subdiffusion coexists with a one-step decay of $F_s$.

Both models of CIL display similar subdiffusive crossovers and reach a diffusive regime in the long time limit. As diffusion coincides with particle exchanges, one might expect that both models are controlled by the network remodeling time, i.e. that their overall structural memory would be controlled by the $\tau_b$. As both systems decorrelate in a single step, we can summarize their structural memory via the relaxation time $F_s(\tau_{\alpha})=1/e$. In Fig.~\ref{fig:tau_chi} (a), we plot $\tau_{\alpha}$ versus $k_{\rm off}$ for both CIL-C and CIL-P. In the case of CIL-P, as the system approaches the elastic network limit, $\tau_\alpha \rightarrow \tau_b$. However, while CIL-C also slows down with decreasing $k_{\rm off}$, $\tau_{\alpha} \ll \tau_{b}$ and shows a nonlinear relationship. We hypothesize that the difference in their behavior might arise from differences in their collective motion.

We quantify collective motion through the dynamical heterogeneity, i.e., the fluctuations of $F_s$ between time origins, measured by the four-point susceptibility~\cite{chi4:LacevicJCP2003,chi4:ToninelliPRE2005}
\begin{equation}
\chi_4(\tau) = N\left[ \langle F_s^2(q,\tau;t) \rangle_t - \langle F_s(q,\tau;t) \rangle_t^2 \right],
\end{equation}
where $F_s(q,\tau;t)$ is $F_s$ for a single time origin $t$, i.e., averaged over particles and directions of $\mathbf{q}$ but not over $t$. In Fig.~\ref{fig:tau_chi} (b) and (c) we plot $\chi_4$ for CIL-C and CIL-P respectively. As is common in glassy materials, $\chi_4$ grows over increasing lag-times until reaching a peak heterogeneity at a lag-time $\tau^*\sim\tau_{\alpha}$. The peak heterogeneity value $\chi_4^*=\chi_4(\tau^*)$ then reports on extent of collective the motion over the structural memory time span. In CIL-P, the similarity of $\tau_{\alpha}$ and $\tau_b$ agrees with the lack of growth in $\chi_4^*$. However, CIL-C shows strong evidence of a growing collective length scale as $\chi_4^*$ continues to increase with decreasing $k_{\rm off}$.

\begin{figure}[t]
\centering
\includegraphics{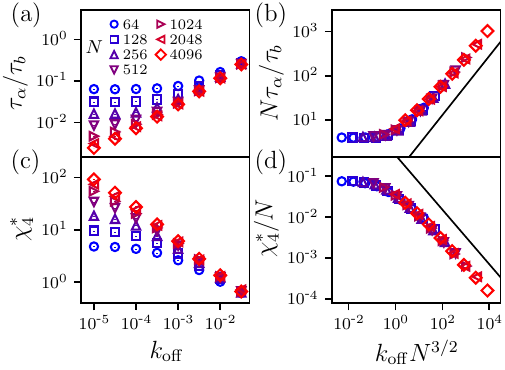}
\caption{\textbf{Slowing bonds, faster relative relaxation.} (a) The ratio $\tau_\alpha/\tau_b$ of the $F_s$ decorrelation time to the bond lifetime vs $k_{\rm off}$ for increasing system sizes from blue to red. (b) Finite-size collapse of $N\tau_\alpha/\tau_b$ vs $k_{\rm off} N^{3/2}$. Black solid line indicates a slope of $2/3$. (c) The peak value $\chi_4^\ast$ of $\chi_4$ vs $k_{\rm off}$. (d) Proposed finite-size collapse of $\chi_4^\ast/N$ vs $k_{\rm off} N^{3/2}$. Black solid line indicates a slope of $-2/3$.}
\label{fig:finite_size_collapse_C}
\end{figure}

As correlation lengths are limited by the finite-sizes of simulations, we next explore how $\tau_{\alpha}$ and $\chi_4^*$ respond to $k_{\rm off}$ as the system size $N$ is varied. In Fig.~\ref{fig:finite_size_collapse_C} (a), we plot $\tau_\alpha/\tau_b$ versus $k_{\rm off}$ for CIL-C with $N = 64$–$4096$ as the ratio reports on how the structural memory diverges from the network's memory of connectivity. $\tau_{\alpha}/\tau_b$ decreases as $k_{\rm off}$ decreases until it reaches a plateau, where the crossover shifts to smaller $k_{\rm off}$ as $N$ increases. In Fig.~\ref{fig:finite_size_collapse_C} (b), we show that the data for all $N$ collapse when $N\tau_\alpha/\tau_b$ is plotted versus $k_{\rm off}N^{3/2}$. For $k_{\rm off}N^{3/2} \lesssim 1$, $N\tau_\alpha/\tau_b$ is constant, and for $k_{\rm off}N^{3/2} \gtrsim 1$, $N\tau_\alpha/\tau_b \propto (k_{\rm off}N^{3/2})^{2/3}$, i.e., $\tau_\alpha/\tau_b \propto k_{\rm off}^{2/3}$ independent of $N$. The dynamical heterogeneity shows the inverse trend. In Fig.~\ref{fig:finite_size_collapse_C} (c), $\chi_4^*$ grows as $k_{\rm off}$ decreases, and in Fig.~\ref{fig:finite_size_collapse_C} (d), $\chi_4^*/N$ collapses on the same variable, with a plateau for $k_{\rm off}N^{3/2} \lesssim 1$ and $\chi_4^* \propto k_{\rm off}^{-2/3}$ above it. These collapses indicate that a single crossover length, $\ell^* \propto k_{\rm off}^{-1/3}$, sets both $\tau_\alpha$ and $\chi_4^*$: for $L < \ell^*$ the size of the correlated regions is set by the box, and for $L > \ell^*$ it is set by $k_{\rm off}$. In CIL-P, finite-size effects are weaker and $\tau_\alpha$ does not collapse on any single variable.  (See Appendix~\ref{appendix:cil_p_finite}).

Previous work on active solids has shown that persistent random forces on an elastic network drive collective flows whose extent grows with the persistence time of the forces~\cite{active_solid:HenkesNatComms2020}. These flows appear in the spatial spectrum of the instantaneous cell velocities: independent forces on each cell give a spectrum that rises towards $q^{-2}$ at small $q$, while forces between pairs of cells give a flat one. In Appendix~\ref{appendix:active_solid}, we measure this instantaneous velocity spectrum for both models. In CIL-C, the small-$q$ spectrum steepens towards $q^{-2}$ as $k_{\mathrm{off}}$ decreases, so at each instant its cells move in increasingly long-wavelength flows. CIL-P, whose forces all act along bonds, has a nearly flat spectrum at every $k_{\mathrm{off}}$. The analogy holds only for the instantaneous velocities, not yet for the displacements over $\tau_\alpha$ that set $\chi_4^*$ and $\tau_\alpha$, but it suggests that an analytical treatment of the tensioned fluid may build on active-solid theory. 

Thus, as $k_{\rm off} \to 0$ and the network approaches a tensioned solid, $\tau_\alpha$ grows as $\tau_b^{1/3}$, so that positions relax ever faster relative to bond turnover. Similarly, $\chi_4^* \propto \tau_b^{2/3}$. This scaling suggests that as the bond dynamics slow, the correlated regions grow and drive faster relaxations relative to the bond dynamics, the opposite of glass-forming liquids, where growing dynamical heterogeneity accompanies slowing relaxation.

In summary, both activity models generate diffusive fluids, with intermediate subdiffusive crossovers, whose $F_s$ decays in a single step without the caging plateau of glass-forming liquids. Both also display collective motion that grows as $k_{\rm off}$ decreases. However, the two models approach the tensioned elastic solid limit at $k_{\rm off} \to 0$ differently, because their collective motion is organized differently across length scales. In CIL-P, the displacement is scale-free, spread equally over every wavelength, and structural relaxation follows bond turnover, $\tau_\alpha \sim \tau_b$: cells forget their positions as their bonds are replaced. In CIL-C, the displacement concentrates at the longest wavelengths, in correlated regions that grow as $\chi_4^* \propto \tau_b^{2/3}$, while relaxation slows only as $\tau_\alpha \propto \tau_b^{1/3}$, so that cells forget their positions long before they forget their bonds. 

\subsection{\label{sec:structure}Long-range Structure in the hysteretic particle model under CIL-C and CIL-P activity}

Both CIL rules direct activity away from a particle's neighbors and towards empty space, a choice we made to keep the tissue locally uniform. However, we have just shown that the two models have very different degrees of heterogeneity and different structural memories. Because density changes through displacement, these differences in motion should leave a mark on the overall structure. To measure structure across wavelengths, we render each configuration as a binary image $I$ (1 inside particles, 0 outside), compute its Fourier transform $\tilde{I}(\mathbf{q})$, and find its spectral density
\begin{equation}
\chi_V(q) = \frac{\left\langle \big| \tilde{I}(\mathbf{q}) \big|^2 \right\rangle}{N\langle a^2 \rangle},
\end{equation}
where $\langle a^2\rangle$ is the mean squared particle area and the average runs over configurations and over wave vectors of magnitude $q$~\cite{hyper:TorquatoPhysReps2018}. $\chi_V$ is an area-weighted structure factor, which we use instead of $S(q)$ because our particles are polydisperse.

\begin{figure}[t]
\centering
\includegraphics{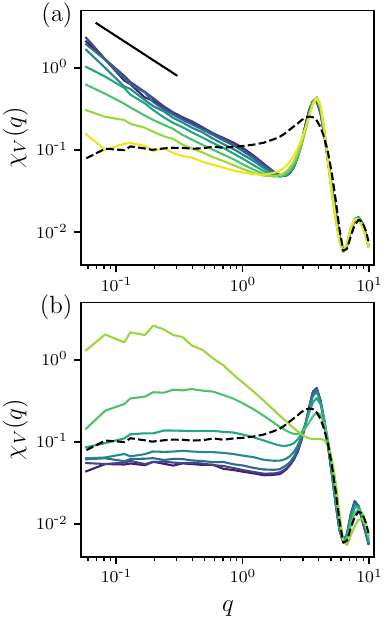}
\caption{\textbf{Density fluctuations are small at the scale of a few particles, $q \sim 1$, and increase slightly (CIL-P) or a lot (CIL-C) at longer length scales ($q \ll 1$).} The area weighted structure factor $\chi_V$ vs wave vector $q$ for (a) CIL-C and (b) CIL-P. $k_{\rm off}$ increases from purple to yellow. RSA is provided for comparison (black dashed line). The black line indicates a slope of $-1$.}
\label{fig:hyperuniformity}
\end{figure}

We plot $\chi_V(q)$ for CIL-C and CIL-P in Fig.~\ref{fig:hyperuniformity}(a) and (b), respectively, together with random non-overlapping disks generated by random sequential addition (RSA) as a reference. The first peak in $\chi_V$ is set by the typical particle spacing. At wavelengths just longer than this spacing, both CIL-C and CIL-P are significantly more uniform than RSA. At small $q$, the slope of $\chi_V$ classifies the large-scale uniformity. For example, a positive slope indicates a hyperuniform system, whose density fluctuations vanish at long wavelengths. In CIL-C, however, we find that the slope approaches $-1$ as $k_{\rm off}$ decreases, so CIL-C is anti-hyperuniform. Although CIL-C is highly uniform locally, its density fluctuations grow continually with wavelength. This anti-hyperuniformity is likely related to the long-wavelength enrichment of the velocity spectrum. In contrast, just as the CIL-P velocity spectrum is scale free, $\chi_V$ for CIL-P is flat. CIL-P is neither hyperuniform nor anti-hyperuniform, although at low $k_{\rm off}$ it remains more uniform than RSA even at the longest wavelengths.

Do these different routes to structural relaxation lead to different mechanical responses? We next measure how each model relaxes stress.

\subsection{\label{sec:mechanics}Mechanics in the hysteretic particle model under CIL-C and CIL-P}

\begin{figure}[t]
\centering
\includegraphics{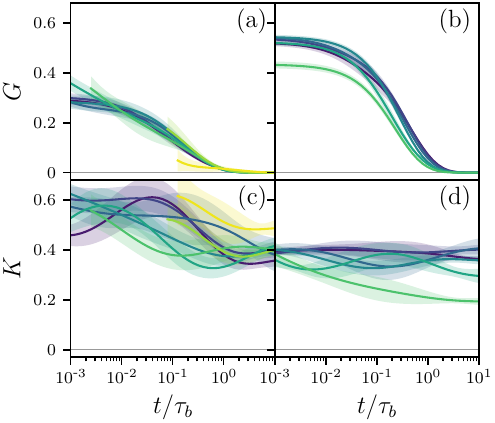}
\caption{\textbf{Linear rheology of the two hysteretic models} Shear modulus $G(t)$ for (a) CIL-C and (b) CIL-P, versus $t/\tau_b$. Bulk relaxation modulus $K(t)$ (c) CIL-C and (d) CIL-P, colored by $k_{\rm off}$ increasing from purple to yellow. Shaded bands show $\pm 1$ standard error (jackknife over 15 pairs of runs). $N=256$.}
\label{fig:rheology}
\end{figure}

The mechanical response of a material can be measured by either applying strain and measuring the resulting stress, or by applying stress and measuring the resulting strain. While a configuration-based definition of active stress has been developed~\cite{swim_pressure:TakatoriPRL2014,swim_pressure:YangSoftMat2014}, it does not necessarily correspond to response-based measurements for active systems that respond to their local environment as in CIL-C~\cite{swim_pressure:SolonNatPhys2015,swim_pressure:FilyJPhysA2017,swim_pressure:EpsteinJCP2019}. Therefore, we measure the effective shear and bulk moduli by applying a weak sinusoidal body force and measuring the displacement it produces~\cite{viscosity:HessJCP2002}.

After equilibration, each cell feels a force $F_0\sin[q(y_i-Y)]$ along $\hat{x}$ (transverse probe) or $\hat{y}$ (longitudinal probe), where $q=2\pi/L$ and $Y$ is the $y$ coordinate of the center of mass. The force drives a displacement mode whose amplitude $U(t)$ is the time integral of the cell velocities along the force, projected onto $\sin[q(y_i-Y)]$. Projecting Eq.~(1) onto the same profile gives, per cell,
\begin{equation}
\gamma\dot{U}(t) = F_0 + \Phi(t),
\label{eq:probe}
\end{equation}
where $\Phi$ is the projection of the internal (pair and active) forces. In linear response, these forces resist the mode with a memory,
\begin{equation}
\Phi(t) = -\frac{q^2}{\rho}\int_0^t G(t-t')\,\dot{U}(t')\,dt',
\label{eq:memory}
\end{equation}
with $\rho=N/L^2$. For the transverse probe, $G$ is the shear relaxation modulus, and its steady flow gives the viscosity $\eta=\int_0^\infty G\,dt$. For the longitudinal probe, $G$ is replaced by the longitudinal modulus $M=K+G$, and we obtain $K$ with $G$ held at its transverse value. We represent each modulus by a discrete relaxation spectrum, a sum of exponentials with fixed, logarithmically spaced relaxation times (a generalized Maxwell model~\cite{viscoelasticity:Ferry1980}), which makes Eq.~(\ref{eq:memory}) linear in the weights, and fit the weights by least squares. To reduce noise, we run each probe at $+F_0$ and $-F_0$ from the same state and average the difference. We use a small forcing of $F_0=0.05\,q$ to sample in the linear regime.

In Fig.~\ref{fig:rheology} (a), (b) we plot the shear response $G$ against $t/\tau_b$ for CIL-C and CIL-P, respectively. We find that both systems follow an approximately one-step decay and fully relax their shear stress by $t \approx \tau_b$, and that the terminal decay collapses on $t/\tau_b$: shear stress is relaxed by the stochastic breaking of bonds. However, CIL-C has a lower plateau in $G$, resulting in a shear viscosity that is approximately half that of CIL-P. In Fig.~\ref{fig:rheology} (c), (d) we plot the bulk response $K$ against $t/\tau_b$ for CIL-C and CIL-P, respectively. In contrast to $G$, $K$ remains positive at all times and does not relax to zero: both materials lose their shear rigidity but retain a bulk modulus, as expected for a liquid.

\begin{figure}[t]
\centering
\includegraphics{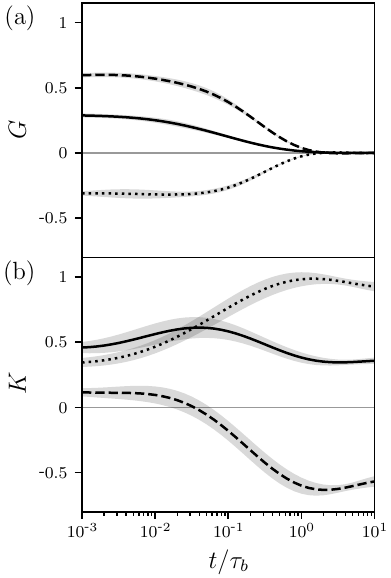}
\caption{\textbf{Crawling activity softens shear but stabilizes bulk.} Decomposing the total (solid lines) CIL-C (a) shear and (b) bulk response into the separate contributions from the pairwise potential (dashed) and active (dotted) for $k_{\rm off}=10^{-5}$. Shaded bands show $\pm 1$ standard error (jackknife over 15 pairs of runs). $N=256$.}
\label{fig:rheology_split}
\end{figure}

To understand the CIL-C mechanical response, we separate the contributions of the pair potential forces and the active crawling forces. Because the probe measures forces, projecting each separately, $\Phi=\Phi_p+\Phi_c$, and fitting each with Eq.~\ref{eq:memory} gives $G=G_p+G_c$ and $K=K_p+K_c$. In Fig.~\ref{fig:rheology_split} we plot this decomposition for CIL-C at the lowest $k_{\mathrm{off}}$. The pair contribution to the CIL-C shear modulus is close to the shear modulus of CIL-P (Fig~\ref{fig:rheology}(b), but the active crawling contribution is negative, canceling about half of the pair contribution, lowering the shear viscosity. Thus, the lower shear modulus and viscosity of CIL-C come from the active crawling, not from a softer network. One possible mechanism for this softening is that shear tilts the bond vectors between neighboring cells, and the crawling activity, which points opposite to the bonds, tilts with the shear and partly cancels their restoring force. The bulk response is more striking. The pair contribution decreases monotonically and becomes negative; the network alone would lose tension in proportion to the area it covers. The active contribution instead rises as bonds turn over and more than compensates, so that the total $K$ remains positive. CIL-C therefore resists changes in density through its activity, consistent with the active crawling opposing the clustering instability of the tensioned network.

\subsection{\label{sec:memory}Crawling activity breaks the Maxwell relation; pulling does not}

\begin{figure}[t]
\centering
\includegraphics{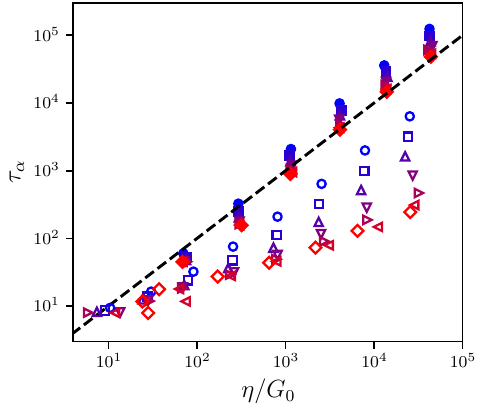}
\caption{\textbf{CIL-P behaves as a simple Maxwell fluid, while in CIL-C the stresses are much longer-lived than the structure.} $F_s$ relaxation time $\tau_\alpha$ vs $\eta/G_0$ relaxation time $\tau_\eta$ for CIL-C (unfilled markers) and CIL-P (filled markers). System size $N$ increases from blue to red. Black dashed line indicates $\tau_\alpha = \eta/G_0$.}
\label{fig:memory}
\end{figure}

The structural and mechanical properties of any material can be thought of in terms of memory: how long it takes particles to reconfigure their positions and how long the material retains patterns of stress. In a Maxwell fluid, a single relaxation time sets both the loss of position memory and the loss of stress memory~\cite{maxwell_relation:DyreRMP2006}. The memory of stress is the relaxation modulus $G(t)=G_0\,e^{-t/\tau}$, which forgets on the time $\tau$, so the viscosity $\eta=\int_0^\infty G(t)\,dt=G\tau$ records how long the stress is remembered, $\tau=\eta/G$. The memory of position is $F_s$, which forgets on $\tau_\alpha$. A Maxwell fluid therefore has $\tau_\alpha\approx\eta/G$. In Fig.~\ref{fig:memory} we test this relation by plotting $\tau_\alpha$ against $\eta/G_0$, where $G_0$ is the plateau modulus of each system.

We find that for all system sizes, CIL-P follows the Maxwell fluid relation. CIL-P is a highly connected tensioned network, and yet through its activity, rearranges and flows like a simple fluid. CIL-C behaves differently. As $k_{\rm off}$ decreases, CIL-C departs from the Maxwell fluid relationship, with the difference growing larger as the system size is increased. The departure from the simple fluid behavior is due to the growing long-range collective motion in CIL-C as $k_{\rm off}$ decreases as where the viscosity grows at a constant rate with $k_{\rm off}$.
Through long-wavelength collective motion, CIL-C fully decorrelates its positions far faster than its stress-carrying bonds: it remembers its stress even when the positions have fully reconfigured and lost all memory of their initial state.

\section{\label{sec:conclusions}Conclusions and Future Directions}

Here, we have characterized the dynamical, structural, and mechanical properties of a sparse fluid under tension -- a uniform, tensioned material  with gaps between particles that is nonetheless able to rearrange. Such a material is possible if activity is supplied to oppose the clustering instability inherent in sparse attractive interparticle interactions, an activity reminiscent of CIL in cells.

However, if both the activity and the attraction depend on particle separations with no stochasticity or time delay, as in the simplest attractive particle model, the system can fall into an absorbing state in which no global flow occurs. 

We avoid the absorbing state with a hysteretic sticking potential, which decouples unbinding from separation. We show that CIL-like behavior can be modeled with either monopole or dipole activity, and that the two lead to distinct material properties. The monopole-driven CIL-C model generates long-wavelength collective displacements that decorrelate positions long before stresses relax, whereas the scale-free collective motion of the dipole-driven CIL-P model decorrelates positions only on the bond lifetime. These different motions leave different large-scale structures: CIL-C is anti-hyperuniform, with density fluctuations that grow with wavelength, while CIL-P is neither hyperuniform nor anti-hyperuniform, and both are more uniform than random packings at smaller scales (i.e. the scale of a few particles). In both models, however, the complete relaxation of stress is carried by the bonds, regardless of how much the positions have decorrelated, and the tension carried by the bonds never relaxes.

Our results also provide metrics for identifying and studying fluids under tension in other systems. In particular, they may help determine which CIL-like mechanism sparse mesenchymal cells use: a crawling force, as in CIL-C, or a pulling force, as in CIL-P. The same comparison could apply to other systems with similar features, such as actin networks~\cite{networks:WangPRXLife2026}, or to bioinspired colloids built with the same interactions. Our models could also be made more realistic, for example with finite CIL repolarization times and noise in the active forces, to test which of these signatures are robust.

Finally, our results suggest ways to incorporate local feedback rules, and with them incorporate design, function, and learning, into these networks. For example, the unbinding rate could respond to the local bond tension, as for catch or slip bonds, or the binding rate could depend on the distance between particles. Each of these rules is a knob that feedback on local or global goals could tune, to design not only a fluid under tension but a rearranging network that can learn and carry out functions, where the memory is not encoded in a specific bond-level property, but in the dynamical node-level properties that generates the rearranging network.

\begin{acknowledgments}
We thank Alessandro Mongera and Gabriella Shih Ping Hsia for scientific discussions and insight into the chick PSM that inspired these models, as well as the experimental image in Fig.~\ref{fig:models}. MLM and ATG acknowledge support from grant number 2023‐329572 from the Chan Zuckerberg Initiative DAF, an advised fund of Silicon Valley Community Foundation, NIH 1R01GM165934, and Syracuse University's Orange Grid research computing cluster. MLM additionally recognizes support for this work from DMR-CMMT-2532170, and conversations at programs at the Kavli Institute for Theoretical Physics (KITP), supported by NSF PHY-2309135 and the Gordon and Betty Moore Foundation Grant No. 2919.02.    
\end{acknowledgments}

\appendix

\section{Attractive particle time-step dependence}
\label{appendix:dt}

\begin{figure}[t]
\centering
\includegraphics{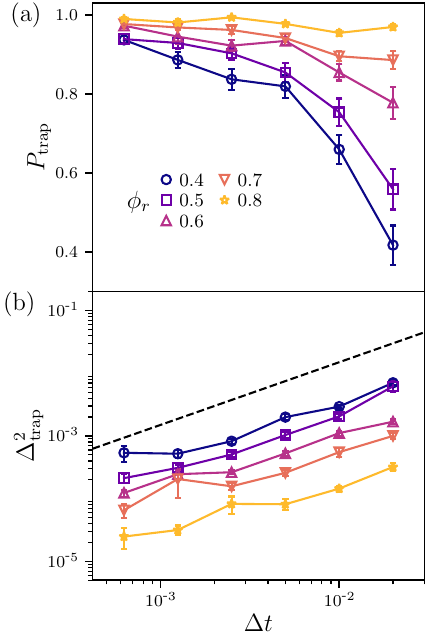}
\caption{\textbf{Integration accuracy effects for the attractive particle model.} (a) The probability of trapping $P_{\rm trap}$ vs the time-step $\Delta t$ used to integrate the simulation. (b) The displacement jitter within a trap vs $\Delta t$. The dashed line indicates slope of 1. The repulsive packing fraction $\phi_r$ increases from purple to orange. Error bars indicate the standard deviation.}
\label{fig:dt}
\end{figure}

In Sec.~\ref{sec:sticky} we presented evidence that the attractive particle model under CIL-C activity evolves towards an absorbing state in which, while every particle is still active, the overall system is trapped and no longer moving. As the simulation has several discontinuities, we present the integration time step $\Delta t$ dependence. In Fig.~\ref{fig:dt} (a), we plot the probability to be trapped over the entire simulation and as $\Delta t$ decreases, the probability of trapping converges. Additionally, the $\Delta t$ effect decreases with increasing repulsive packing fraction $\phi_r$. In Fig.~\ref{fig:dt} (b) we plot the average local displacements of the system when it is trapped $\Delta^2_{\rm trap}$ versus $\Delta t$. We find that the increase in $P_{\rm trap}$ due to both $\Delta t$ and $\phi_r$ corresponds to decreasing $\Delta^2_{\rm trap}$. In other words, as the time step decreases and as the attractive shell shrinks, the local jitter of the system decreases, leading to a decreased chance for the system to be knocked out of the stable absorbed state. Therefore, we expect that it is possible to reduce the probability of finding the absorbing state in the attractive particle model, e.g. by adding an activity repolarization time or rotation noise at the expense of opposing the clustering instability.

\begin{figure}[t]
\centering
\includegraphics{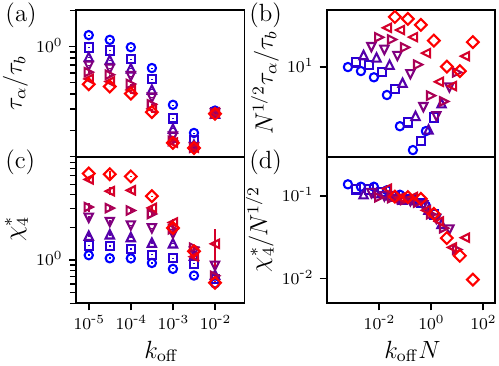}
\caption{\textbf{Finite-size scaling for CIL-P.} (a) Ratio $\tau_\alpha/\tau_b$ of the $F_s$ relaxation time to the bond lifetime versus $k_{\rm off}$ for $N = 64$–$4096$ (blue to red). Points at $k_{\rm off} = 10^{-2}$ lie outside the tensioned-fluid phase. (b) $N^{1/2}\tau_\alpha/\tau_b$ versus $k_{\rm off}N$: the variable that collapses $\chi_4^*$ in (d) does not collapse $\tau_\alpha$. (c) Peak value $\chi_4^*$ of $\chi_4$ versus $k_{\rm off}$. (d) Finite-size collapse of $\chi_4^*/N^{1/2}$ versus $k_{\rm off}N$.}
\label{fig:finite_size_collapse_P}
\end{figure}

\section{CIL-P finite size}
\label{appendix:cil_p_finite}
In Fig.~\ref{fig:finite_size_collapse_P}(a), we show that for the CIL-P model, $\tau_\alpha/\tau_b$ {\it increases} as $k_{\rm off}$ decreases, reaching $\tau_\alpha/\tau_b \approx 0.5$–$1$ at the lowest $k_{\rm off}$. This trend suggests that positional relaxation in CIL-P is set largely by bond turnover. At fixed $k_{\rm off}$, $\tau_\alpha/\tau_b$ still decreases with $N$, but weakly, by a factor of $\approx 2.5$ from $N = 64$ to $4096$, compared with a factor of $\approx 60$ in CIL-C. However, $\tau_\alpha \sim \tau_b$ does not mean that CIL-P lacks collective motion. In Fig.~\ref{fig:finite_size_collapse_P}(c), we find that $\chi_4^*$ also grows as $k_{\rm off}$ decreases, although it grows much more slowly with $N$ than in CIL-C. In Fig.~\ref{fig:finite_size_collapse_P}(d), we show that $\chi_4^*/N^{1/2}$ collapses when plotted versus $k_{\rm off}N$, with a plateau for $k_{\rm off}N \lesssim 1$, so that $\chi_4^* \propto N^{1/2}$ in small systems rather than $\chi_4^* \propto N$ as in CIL-C. In contrast to CIL-C, the variable that collapses $\chi_4^*$ does not collapse $\tau_\alpha$. In Fig.~\ref{fig:finite_size_collapse_P}(b), $N^{1/2}\tau_\alpha/\tau_b$ plotted versus $k_{\rm off}N$ separates by $N$. Thus, in CIL-P the correlated regions grow slightly as bonds become longer lived, but their growth does not set the relaxation time.

\section{Active solid velocity spectrum}
\label{appendix:active_solid}

\begin{figure}[t]
\centering
\includegraphics{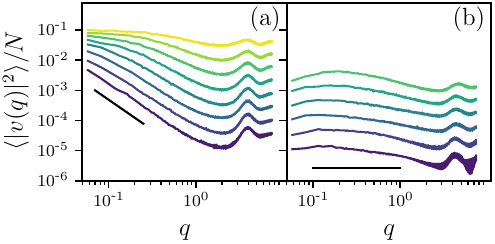}
\caption{\textbf{Crawling drives long-wavelength flows; pulling does not.} Velocity spectrum $\langle|v(q)|^2\rangle/N$ against $q$ for (a) CIL-C and (b) CIL-P at $N=4096$. Colors indicate $k_{\rm off}$ increasing from purple to yellow. Black lines show slopes $-2$ (a) and $0$ (b).}
\label{fig:v_spectrum}
\end{figure}

Is the collective motion of CIL-C that of an elastic solid driven by persistent random forces? To test this picture, we measure the velocity spectrum of Henkes \textit{et al.}~\cite{active_solid:HenkesNatComms2020}, $\langle|v(q)|^2\rangle/N$ with $\mathbf{v}(\mathbf{q}) = \sum_j \mathbf{v}_j e^{i\mathbf{q}\cdot\mathbf{r}_j}$, where $\mathbf{v}_j$ is the instantaneous velocity of cell $j$, averaged over wave vectors in each shell of $|\mathbf{q}|$ and over time. In linear response, each polarization $X$ gives
\begin{equation}
\frac{\langle|v_X(q)|^2\rangle}{N} = \frac{S_X(q)/\gamma^2}{1 + \xi_X^2 q^2}, \qquad \xi_X^2 = \frac{M_X\tau_p}{\rho\gamma},
\label{eq:henkes}
\end{equation}
where $S_X$ is the spectrum of the active force, $\tau_p$ its correlation time, $\gamma$ the friction per cell and $M_X$ the shear or longitudinal modulus. Independent forces on each cell ($S_X = f_0^2/2$) give a flat spectrum that falls as $q^{-2}$ for $q\xi \gg 1$. Forces between pairs of cells instead cancel at long wavelengths: with $\mathbf{f}_{ij} = -\mathbf{f}_{ji}$, the force on mode $\mathbf{q}$ is $\sum_{i<j}\mathbf{f}_{ij}\,(e^{i\mathbf{q}\cdot\mathbf{r}_i} - e^{i\mathbf{q}\cdot\mathbf{r}_j}) \approx \sum_{i<j} i(\mathbf{q}\cdot\mathbf{r}_{ij})\,\mathbf{f}_{ij}\,e^{i\mathbf{q}\cdot\mathbf{r}_i}$ for $q r_{ij} \ll 1$, which vanishes as $q$. Pair forces therefore act as force dipoles with $S_X \propto q^2$, and this factor cancels the $q^{-2}$ decay to give a flat spectrum for $q\xi \gg 1$.

In Fig.~\ref{fig:v_spectrum} (a) and (b), we plot the instantaneous velocity spectrum for CIL-C and CIL-P respectively. For CIL-C, we find that as $k_{\rm off}$ decreases, the small $q$ slope steepens from relatively flat towards $-2$ in the same manner as predicted for an active solid. The similarity between active solid theory and CIL-C suggests that, at least at short times, while the unbinding events drive the system to flow like a fluid, the unbinding events trigger decorrelations of the activity on a highly coordinated elastic network and drive collective motions via a similar mechanism. For CIL-P, we find that the velocity spectrum is relatively flat over all $q$, as expected for dipolar driving. We leave further analytic treatment across longer time lags for future work.

\noindent
\textbf{Code Availability.} The discrete element simulation code developed here is available at: \url{https://github.com/agrigas115/hysteretic_sticking_plus_CIL}.

\end{document}